\documentclass[aps,prc,twocolumn,amsmath,amssymb,floatfix,superscriptaddress]
{revtex4-1}
\usepackage{mathptmx}
\usepackage[T1]{fontenc}
\usepackage{CJKutf8}
\usepackage{graphicx}
\usepackage{bm}
\usepackage{dcolumn}
\usepackage{multirow}
\usepackage{xcolor}
\usepackage{epstopdf}
\usepackage{booktabs}

\usepackage[
breaklinks,pdfstartview=FitH,CJKbookmarks=true,
bookmarksnumbered=true,bookmarksopen=true,pdfborder={0 0 1},
colorlinks=true,linkcolor=blue,urlcolor=blue,anchorcolor=blue,citecolor=blue
            ]{hyperref}

\makeatletter
\begin{document}

\begin{CJK*}{UTF8}{gbsn}
\title{
Hypernuclear structure with the new leading order covariant chiral hyperon-nucleon force
}

\author{Xuan Zeng}
\affiliation{
Department of Physics, East China Normal University,
Shanghai 200241, China}

\author{Ru-You Zheng}
\author{Zhi-Wei Liu}
\affiliation{School of Physics, Beihang University, Beijing 102206, China}

 \author{Li-Sheng Geng}\email{lisheng.geng@buaa.edu.cn}
\affiliation{School of Physics, Beihang University, Beijing 102206, China}
\affiliation{Centrale Pékin, Beihang University, Beijing, 100191, China}
\affiliation{Peng Huanwu Collaborative Center for Research and Education, Beihang University, Beijing 100191, China}
\affiliation{Beijing Key Laboratory of Advanced Nuclear Materials and Physics, Beihang University, Beijing 102206, China }
\affiliation{Southern Center for Nuclear-Science Theory (SCNT), Institute of Modern Physics, Chinese Academy of Sciences, Huizhou 516000, China}

\author{Xian-Rong~Zhou} \email{xrzhou@phy.ecnu.edu.cn}
\affiliation{
Department of Physics, East China Normal University,
Shanghai 200241, China}

\date{\today}

\begin{abstract}
We have studied single-$\Lambda$ finite hypernuclear systems spanning from light to heavy masses, employing a new microscopic $\Lambda N$ interaction derived from in-medium interactions within relativistic Brueckner-Hartree-Fock calculations using the leading-order covariant chiral hyperon-nucleon force. Without any adjustable parameters, we have successfully reproduced the experimental results for the single-$\Lambda$ binding energies. Although some discrepancies persist in light hypernuclei, the overall calculated binding energies show excellent agreement with the experimental data, and outperforming other microscopic interactions. This study further demonstrates the validity of the leading-order covariant chiral hyperon-nucleon potential and provides a practical set of microscopic interactions for the Skyrme-Hartree-Fock framework.
\end{abstract}

\maketitle

\end{CJK*}

\section{Introduction}
\label{intro}

The properties of hypernuclei and hypermatter are highly sensitive to hyperon-nucleon ($YN$) and hyperon-hyperon ($YY$) interactions \cite{gai2016,youxianhe2021,ns2020,expyn1,expyn2}. On the one hand, the properties of hypernuclei and hypermatter impose important constraints on $YN$ and $YY$ interactions \cite{gai2023,hyptest2023,yn2,ns2013}. Among the relevant systems, finite nuclei and neutron stars are two classical platforms used to constrain $YN$ and $YY$ interactions. On the other hand, improved knowledge of $YN$ and $YY$ interactions is essential for exploring the properties of strange hadronic matter \cite{hanszong,midu2015,zheng2025}. The hyperon puzzle in neutron stars drives the demand for more precise $YN$ interactions. Theoretical predictions indicate that the softening effect caused by hyperons reduces the maximum mass of neutron stars, but observations confirm the existence of 2$M_{\odot}$ neutron stars \cite{ns2020,ns2013,ns2015,nsdata2016,nsab2024}. This discrepancy indicates the necessity of introducing repulsive components at high densities in order to account for hypernuclear data \cite{gai2023,nsab2024}.

Theoretically, $YN$ and $YY$ interactions can be classified into bare interactions and in-medium interactions. Bare interactions can be constructed either through phenomenological models including the Nijmegen and $\mathrm{J\ddot{u}lich}$ meson-exchange potentials \cite{nji1,bare89,nsc97,nji3,ju1,ju2}, quark models based on SU(3) or SU(6) symmetry \cite{quark1,quark2}, or through ab initio approaches including nonrelativistic chiral effective field theory \cite{nonrela1,nonrela2,nonrela3}, relativistic chiral effective field theory which is of current interest for its consistent incorporation of Lorentz symmetry \cite{yn1,yn2,yn3,yn4,yn5} \footnote{We note that a high-precision nucleon-nucleon interaction has been constructed in the covariant framework \cite{gengnn}.}, and lattice quantum chromodynamics (QCD) which allows direct numerical simulations of quantum chromodynamics \cite{qcd1,qcd2,qcd3}.

The in-medium interaction, which refers to the effective interaction incorporating quantum many-body effects, is typically constructed through phenomenological approaches, 
such as relativistic mean-field theory (RMF) \cite{rmf1,rmf2,rmf3,rmf4}, Skyrme-Hartree-Fock methods (SHF) \cite{beChen2022,guo22,xue2023,liu2023,xue2024}, Woods-Saxon potentials incorporating effective hyperon-nucleon interactions \cite{gai1988,gai2023}. 
In-medium interactions can also be constructed by microscopic ab initio methods, which derive the effective interaction from bare interactions without any empirical parameters. The Brueckner-Hartree-Fock (BHF) approach employs the G-matrix formalism to obtain the in-medium effective interaction starting from the bare interaction \cite{hans1988,bhfquark1,bhf2,hans2011,bhf2015,bhf2023}.  The relativistic Brueckner-Hartree-Fock (RBHF) method incorporates relativistic effects and is able to reproduce the empirical saturation properties of nuclear matter accurately using only two-body interactions \cite{zheng2025,rbhf1,rbhf2,rbhf3,rbhf4}.

In 2025, an in-medium $\Lambda$-nucleon ($\Lambda N$) interaction has been obtained within the RBHF framework based on a leading-order (LO) covariant chiral effective field for the first time, in which 12 low-energy constants (LECs), required by Lorentz invariance at LO, are determined by fitting to 36 sets of hyperon-nucleon scattering data \cite{zheng2025}. Despite the remaining cutoff dependence of this interaction, which was constructed using cutoffs ranging from 550 MeV to 700 MeV, it benefits from a more microscopic theoretical foundation and symmetry constraints. The model self-consistently reproduces both the scattering cross sections and the empirical single-particle potential of the $\Lambda$ hyperon in nuclear matter. 
 
Although the LO covariant chiral in-medium $\Lambda N$ interaction yields empirical single-particle potentials in nuclear matter that are consistent with empirical values, it is still restricted to nuclear matter and cannot be directly compared with the richer and more detailed set of data from finite hypernuclei, thus lacking additional constraints from finite nucleus experiments. 
Moreover, as mentioned earlier, the 12 LECs are obtained by fitting 36 existing $\Lambda N$ scattering data sets. Compared with nucleon-nucleon ($NN$) interactions \cite{nn1,nn2,nn3,nn4,nn5}, $\Lambda N$ interactions involve more coupled channels but are constrained by significantly fewer experimental data \cite{expyn1,expyn2,gai2016}, and in contrast to the extensive and detailed measurements of single-$\Lambda$ hypernuclear binding energies and level structures, the available $\Lambda N$ scattering data remain both sparse and imprecise \cite{yys1,yys3,yys8,yys6,yys5,yys7}. 
This provides a natural motivation to extend the LO covariant chiral in-medium $\Lambda N$ interaction to finite hypernuclear systems, enabling direct comparisons with richer and more detailed experimental data to further constrain and validate the $\Lambda N$ interaction in finite nuclear systems.

For finite nuclear systems, the SHF method is an efficient and self-consistent mean-field approach that is not hindered by increasing core mass \cite{liu25,liu26,liu27}. 
In the past, the in-medium interactions derived from the G-matrix-based BHF method have been successfully extended to finite nuclear systems and have achieved considerable success within the SHF framework \cite{hansnsc89,hansnsc97,zhou07,hansesc08}. 
In-medium interactions based on Nijmegen soft-core potentials, including NSC89 \cite{hansnsc89,hans1988,bare89}, NSC97a \cite{nsc97,hansnsc97}, NSC97f \cite{nsc97,hansnsc97,bare97}, and the extended soft-core potential ESC08 \cite{nji3,hansesc08,Rbare08}, have been successfully extended to finite nuclear systems and implemented within the SHF framework. In the SHF framework, by retaining the microscopic foundation without introducing adjustable parameters, this method supports calculations across the full nuclear chart. 

Among these, NSC89 and NSC97f have been shown to reasonably reproduce certain structural features of $\Lambda$ hypernuclei. The ESC08 potential initially exhibited some overbinding in describing single-$\Lambda$ hypernuclei, but its accuracy was significantly improved through the inclusion of correction terms. Similar refinements have also been applied to other interactions, effectively addressing earlier deficiencies and enabling a better description of hypernuclear properties \cite{midufanhan,a7term,a7same,guo22}.
These successes demonstrate the effectiveness and practicality of these interactions. Notably, these microscopic bare potentials are transformed via the G-matrix into microscopic effective interactions used in SHF calculations of hypernuclei, further reinforcing the microscopic basis of the SHF theory.

The motivation of this work is to construct a new microscopic interaction based on the in-medium $\Lambda N$ interaction derived from RBHF calculations of nuclear matter using the leading-order covariant chiral $YN$ force \cite{zheng2025}, and to apply it within the SHF framework for calculations of finite hypernuclei. Due to the cutoff dependence of the leading-order covariant chiral $YN$ force, we adopted the cutoff values of 550 MeV and 700 MeV, which correspond to the upper and lower bounds studied in Ref.~\cite{zheng2025}, to construct our interactions. This approach allows us to systematically evaluate the performance of the interaction in finite hypernuclei. The interactions do not involve adjustable parameters. The framework is further extended to the SHF with Bardeen-Cooper-Schrieffer (BCS) pairing correlations \cite{deform1,zhou07,deltapair2,bcslight,bcsblock1,peterring}, allowing calculations within the range of available single-$\Lambda$ hypernuclear data. 
The results are compared with available experimental data and other microscopic interaction models.

The paper is organized as follows.
In Sec.~\ref{sec2}, a brief introduction is given to the theoretical SHF approach for hypernuclei and the formalism used to derive an effective interaction from microscopic RBHF calculations.
In Sec.~\ref{sec3}, the calculated results of RCh-LO550, RCh-LO700, and selected microscopic interactions are presented, along with comparisons with experimental data.
Finally, a summary is given in Sec.~\ref{sec4}.

\section{Theoretical framework}
\label{sec2}

 We apply the covariant chiral effective $\Lambda N$ interaction derived from the RBHF approach to our SHF mean-field framework in the form of an energy density functional. The SHF equations are solved self-consistently in an axially symmetric coordinate space representation. Within this framework, the total energy of a hypernucleus is given by \cite{Rayet81,hansnsc89,hansnsc97,hansesc08,zhou07,guo22,xue2024}
 \begin{equation}\label{EE}
    E=\int d^3 r \varepsilon(r). 
\end{equation}
The energy density functional can be written as
\begin{equation}
 \varepsilon= \varepsilon_{NN}[\tau_n,\tau_p,\rho_n,\rho_p,\boldsymbol{J}_{n}, \boldsymbol{J}_{p}]+\varepsilon_{N \Lambda}[\tau_{\Lambda},\rho_{\Lambda},\rho_N],
\end{equation}
where $\varepsilon_{NN}$ and $\varepsilon_{N\Lambda}$ denote the energy density functionals corresponding to the $NN$ and $\Lambda N$ interactions, respectively. They depend on the one-body density $\rho_q$, kinetic energy density $\tau_q$, and spin-orbit current $\mathbf{J}_q$
\begin{equation}
    \left[\rho_{q}, \tau_{q}, \boldsymbol{J}_{q}\right]=\sum_{k=1}^{N_{q}} n_{q}^{k}\left[\left|\phi_{q}^{k}\right|^{2},\left|\nabla \phi_{q}^{k}\right|^{2}, \phi_{q}^{k^{*}}\left(\nabla \phi_{q}^{k} \times \boldsymbol{\sigma}\right) / k\right].
\end{equation}
The single-particle wave functions $\phi_q^k$ represent the $k$th occupied states of different particle species $q =\{ n, p, \Lambda$\}, obtained through self-consistent calculations. The occupation probabilities $n_q^k$ are computed within the BCS approximation considering nucleons only. The pairing interaction between nucleons is modeled by a density-dependent $\delta$ force \cite{deltapair1,deltapair2}
\begin{equation}
    V_{q}\left(\boldsymbol{r}_{1}, \boldsymbol{r}_{2}\right)=V_{q}^{\prime}\left[1-\frac{\rho_{N}\left(\left(\boldsymbol{r}_{1}+\boldsymbol{r}_{2}\right) / 2\right)}{0.16 \mathrm{fm}^{-3}}\right] \delta\left(\boldsymbol{r}_{1}-\boldsymbol{r}_{2}\right),
\end{equation}
pairing strengths $V_p = V_n = -410\, \mathrm{MeV\, fm^3}$ are used for light nuclei~\cite{bcslight}, while $V_p = -1146\, \mathrm{MeV\, fm^3}$ and $V_n = -999\, \mathrm{MeV\, fm^3}$ are adopted for medium-mass and heavy nuclei~\cite{zhou07}. A smooth energy cutoff is employed in the BCS calculations~\cite{bcsblock1}. In the case of an odd nucleon number, the orbit occupied by the unpaired nucleon is blocked as described in Ref.~\cite{peterring}.

Regarding the term $\varepsilon_{NN}$, since hyperon observables are largely insensitive to the choice of the $NN$ interaction~\cite{zhou07,beChen2022,Xue22}, we adopt the SLy4 parameterization as the $NN$ interaction, which provides a good description of the binding energies of the light and heavy isotopes~\cite{liu25,liu26,liu27,sly4,sly4hao1}.
Due to the presence of hyperon, this term can be expressed as~\cite{hansnsc89,hansnsc97,hansesc08}
\begin{equation}
    \varepsilon_{\Lambda}=\frac{\tau_{\Lambda}}{2 m_{\Lambda}}+\varepsilon_{N \Lambda}\left(\rho_{N},\rho_{\Lambda}\right).
\end{equation}
we aim to connect the density functional form of the $\Lambda N$ interaction with the results obtained from relativistic RBHF calculations. A detailed discussion can be found in Ref.~\cite{hansnsc89}. The presence of hyperons gives rise to an additional contribution to the binding energy per baryon
\begin{equation}
\begin{aligned}
    \rho_{\Lambda}\frac{B}{A}(0,\rho_{\Lambda})+\varepsilon_{N \Lambda}\left(\rho_{N}, \rho_{\Lambda}\right) &=(\rho_N+\rho_{\Lambda})\frac{B}{A}(\rho_N+\rho_{\Lambda}) \\
     &-\rho_N\frac{B}{A}(\rho_N,0).
     \end{aligned}
\end{equation}
Consequently, the energy density functional can be constructed in terms of the nucleon and hyperon single-particle potentials \cite{hansnsc89}, $U_N(k)$ and $U_{\Lambda}(k)$, obtained from RBHF calculations \cite{zheng2025}
\begin{equation}
    \begin{aligned}\label{eqspotrbhf}
\epsilon_{N \Lambda}\left(\rho_{N}, \rho_{\Lambda}\right)= & 2 \sum_{k<k_{F}^{(\Lambda)}} U_{\Lambda}(k) \\
& +2 \sum_{k<k_{F}^{(N)}}\left[\left.U_{N}^{(N)}(k)\right|_{\rho_{\Lambda}}-\left.U_{N}^{(N)}(k)\right|_{\rho_{\Lambda}=0}\right].
\end{aligned}
\end{equation}
The term $U_N^{(N)}$ represents the single-particle potential of a nucleon induced by the nucleonic medium. Its dependence on the hyperon density is demonstrated to be weak, so the dominant contribution of Eq.(\ref{eqspotrbhf}) arises from the first term of the equation \cite{hansnsc89}. Therefore, in the present work, the effect of the hyperon density on the nucleon single-particle potential in the nucleonic medium is neglected.
\begin{table}
\centering
\caption{Parameter sets of the energy density functional and $\Lambda$ effective mass, RCh-LO550 and RCh-LO700, as given in Eqs.~(\ref{eqinter}) and (\ref{eqmass}). Since only the $\Lambda N$ interaction is considered, the $\Lambda \Lambda$ interaction terms satisfy $\varepsilon_7 - \varepsilon_9 = 0$.}
\label{table1}
\begin{tabular}{lcccccccccc}
\toprule
\toprule
Functional 
& $\varepsilon_1$ & $\varepsilon_2$ & $\varepsilon_3$ & $\varepsilon_4$ 
& $\varepsilon_5$ & $\varepsilon_6$  
& $\mu_1$ & $\mu_2$ & $\mu_3$ & $\mu_4$ \\
\midrule
RCh-LO550    & 359 & 1321 & 750  & 992 & 4738 & 13699 & 1.01 & 3.02 & 5.56 & 3.0 \\
RCh-LO700    & 369 & 1402 & 1894 & 965 & 4021 & 4997  & 1.00 & 2.15 & 5.03 & 6.0 \\
\bottomrule
\bottomrule
\end{tabular}
\end{table}

\begin{figure}[t]
\hspace{-3em} 
\centerline{\includegraphics[scale=0.35]{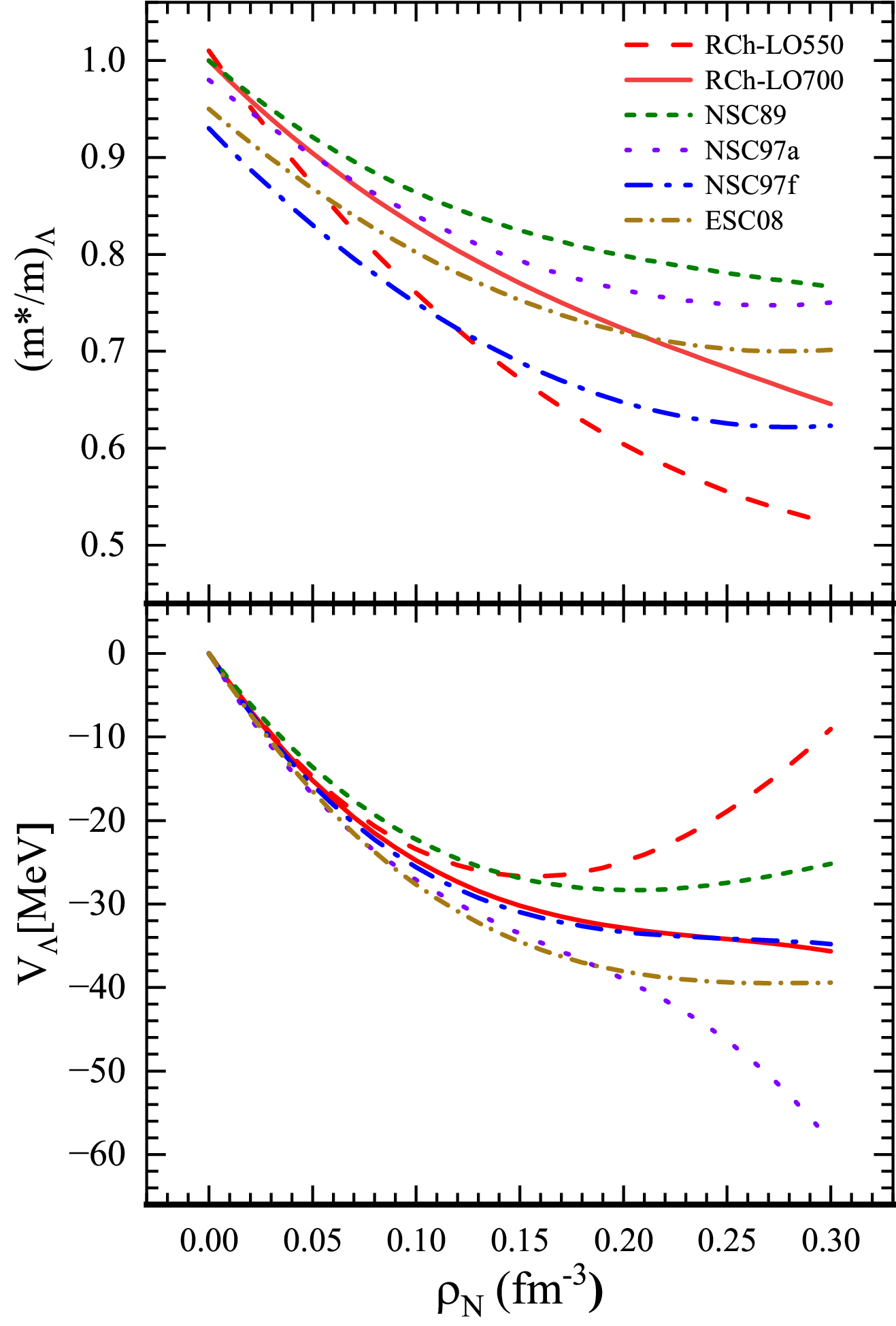}}
\vskip-2mm
\caption{
The effective mass and mean field as functions of nucleon density $\rho_N$ in pure nucleonic matter ($\rho_{\Lambda}=0$). The red dashed and red solid lines represent our parameter sets RCh-LO550 and RCh-LO700, respectively. The green short-dashed line corresponds to the NSC89 interaction, derived by extending the potential to finite nucleus calculations through a combination of the BHF method and the SHF approach \cite{hansnsc89,bare89,hans1988}. The purple dotted line and blue dash-dotted line correspond to the NSC97a and NSC97f parameter sets, respectively, both of which incorporate channels with strangeness $S = -1, -2, -3, -4$ within the NSC97 series \cite{nsc97,hansnsc97}. The yellow short dot-dashed line corresponds to the ESC08 interaction, which is based on the Nijmegen extended soft-core $YN$ and $YY$ potentials and incorporates new meson-exchange and meson-pair exchange interactions \cite{hansesc08,nji3}.
}
\label{fig1}
\end{figure}

In RBHF calculations, the effective mass arising from the correction due to the attractive interaction of scalar meson exchange in a medium can be constructed from the scalar potential \cite{zheng2025,rbhf2}
\begin{equation}
\label{eqsmassrbhf}
    m^*_{\Lambda}=m_{\Lambda}+U^{\Lambda}_S,
\end{equation}
where $U^{\Lambda}_S$ is the $\Lambda$ scalar potential obtained in the RBHF theory with the leading order covariant chiral YN potentials. By introducing an ideal Fermi gas of hyperons, we finally get the energy density functional for hyperons 
\begin{equation}
\begin{aligned}
    \epsilon_{\Lambda}= & \frac{1}{2 m_{\Lambda}} \tau_{\Lambda}+\epsilon_{N \Lambda}\left(\rho_{N}, \rho_{\Lambda}\right) \\
& +\left(\frac{m_{\Lambda}}{m_{\Lambda}^{*}\left(\rho_{N}\right)}-1\right)\left(\frac{\tau_{\Lambda}}{2 m_{\Lambda}}-\frac{3}{5} \frac{\left(3 \pi^{2}\right)^{2 / 3}}{2 m_{\Lambda}} \rho_{\Lambda}^{5 / 3}\right).
\end{aligned}
\end{equation}
Through the variation of the total energy Eq.~(\ref{EE}) one derives the SHF Schr\"odinger equation for both nucleons and hyperons
\begin{equation}
    \left[-\boldsymbol{\nabla} \cdot \frac{1}{2 m_{q}^{*}(\boldsymbol{r})} \boldsymbol{\nabla}+V_{q}(\boldsymbol{r})-i \boldsymbol{W}_{q}(\boldsymbol{r}) \cdot(\boldsymbol{\nabla} \times \boldsymbol{\sigma})\right] \phi_{q}^{k}(\boldsymbol{r})=e_{q}^{k} \phi_{q}^{k}(\boldsymbol{r})\:,
\end{equation}
where $V_{q}(\boldsymbol{r})$ is the density-dependent central part of the mean field for different particles, while $\boldsymbol{W}_{q}$ represents the spin-orbit coupling interaction between nucleons~\cite{shf73,deltapair1}. Since the spin-orbit interaction among hyperons is negligible at the current level of approximation~\cite{nospin}, it is neglected in this work. The mean-field potentials for nucleons $V_N$, and hyperons $V_{\Lambda}$, can be expressed as
\begin{align}\label{potshf}
V_{N}= & V_{N}^{\mathrm{SHF}}+\frac{\partial \epsilon_{N \Lambda}}{\partial \rho_{N}} \nonumber \\
& +\frac{\partial}{\partial \rho_{N}}\left(\frac{m_{\Lambda}}{m_{\Lambda}^{*}\left(\rho_{N}\right)}\right)\left(\frac{\tau_{\Lambda}}{2 m_{\Lambda}}-\frac{3}{5} \frac{\left(3 \pi^{2}\right)^{2 / 3}}{2 m_{\Lambda}} \rho_{\Lambda}^{5 / 3}\right),\\
V_{\Lambda}=&\frac{\partial \epsilon_{N \Lambda}}{\partial \rho_{\Lambda}}-\left(\frac{m_{\Lambda}}{m_{\Lambda}^{*}\left(\rho_{N}\right)}-1\right) \frac{\left(3 \pi^{2}\right)^{2 / 3}}{2 m_{\Lambda}} \rho_{\Lambda}^{2 / 3},
\end{align}
where $V^{\mathrm{SHF}}_{N}$ denotes the Skyrme mean field for nucleons in the absence of hyperons. We parameterize $\varepsilon_{\Lambda N}$ and the effective mass $m_{\Lambda}^*$ following the form given in Ref.~\cite{hansnsc89}, where the densities $\rho_N$ and $\rho_{\Lambda}$ are expressed in units of $\mathrm{fm^{-3}}$, and $\varepsilon_{\Lambda N}$ is in units of $\mathrm{MeV\, fm^{-3}}$
\begin{equation}
    \begin{aligned}\label{eqinter}
\varepsilon_{N \Lambda}= & -\left(\varepsilon_{1}-\varepsilon_{2} \rho_{N}+\varepsilon_{3} \rho_{N}^{2}\right) \rho_{N} \rho_{\Lambda} \\
& +\left(\varepsilon_{4}-\varepsilon_{5} \rho_{N}+\varepsilon_{6} \rho_{N}^{2}\right) \rho_{N} \rho_{\Lambda}^{5 / 3} \\
& -\left(\varepsilon_{7}-\varepsilon_{8} \rho_{\Lambda}+\varepsilon_{9} \rho_{\Lambda}^{2}\right) \rho_{\Lambda}^{2} ,
\end{aligned}
\end{equation}

\begin{equation}\label{eqmass}
    \frac{m^*_{\Lambda}}{m_{\Lambda}}(\rho_N)= -\mu_1-\mu_2\rho_N+\mu_3\rho_N^2-\mu_4\rho_N^3  .
\end{equation}

It should be noted that the nucleon density in our model implicitly incorporates variations in hyperon density. The connection between the energy density functional within the SHF framework and the single-particle potentials from RBHF calculations, together with a phenomenological parameterization of the effective mass, is established to incorporate the two sets of results with cutoffs of $550 \mathrm{MeV}$ and $700 \mathrm{MeV}$ from the covariant chiral effective $\Lambda N$ interaction in Ref.~\cite{zheng2025}. The covariant chiral $YN$ interaction with a cutoff in the range of 550-700 MeV can well reproduce the experimental scattering cross-section results. In nuclear matter calculations, the effective interaction $\Lambda N$ within this cutoff range can also provide a range of single-particle potentials that are in excellent consistency with the empirical values. Two parameter sets for finite hypernuclei calculations within the SHF model are thus derived, hereafter denoted as RCh-LO550 and RCh-LO700, respectively. The specific values of these parameter sets are provided in Table~\ref{table1}.

It is assumed that all nucleons and hyperons move within a fixed background, as the overall motion is physically irrelevant and only relative motion is meaningful. To account for the center-of-mass correction, we replace the bare mass with the following form \cite{shf73,liu25,liu26,liu27,shf1972}
\begin{equation}
    \frac{1}{m_{q}} \rightarrow \frac{1}{m_{q}}-\frac{1}{M},
\end{equation}
where $M = (N_n + N_p) m_N + N_{\Lambda} m_{\Lambda}$ represents the total mass of the hypernucleus.

In our model approximation, the mean field is assumed to be axially symmetric, and the deformed SHF Schr\"odinger equation is solved in cylindrical coordinates $(r, z)$ using a deformed harmonic oscillator basis \cite{shf73,liu25,liu26,liu27}. This allows us to calculate deformed hypernuclei under quadrupole constraints. The optimal quadrupole deformation parameters are determined by minimizing the energy-density functional
\begin{equation}
    \beta_{2}^{(q)} \equiv \sqrt{\frac{\pi}{5}} \frac{\left\langle 2 z^{2}-r^{2}\right\rangle_{q}}{\left\langle r^{2}+z^{2}\right\rangle_{q}}.
\end{equation}

\begin{figure*}[t]
\vskip-2mm
\centerline{\includegraphics[scale=0.45]{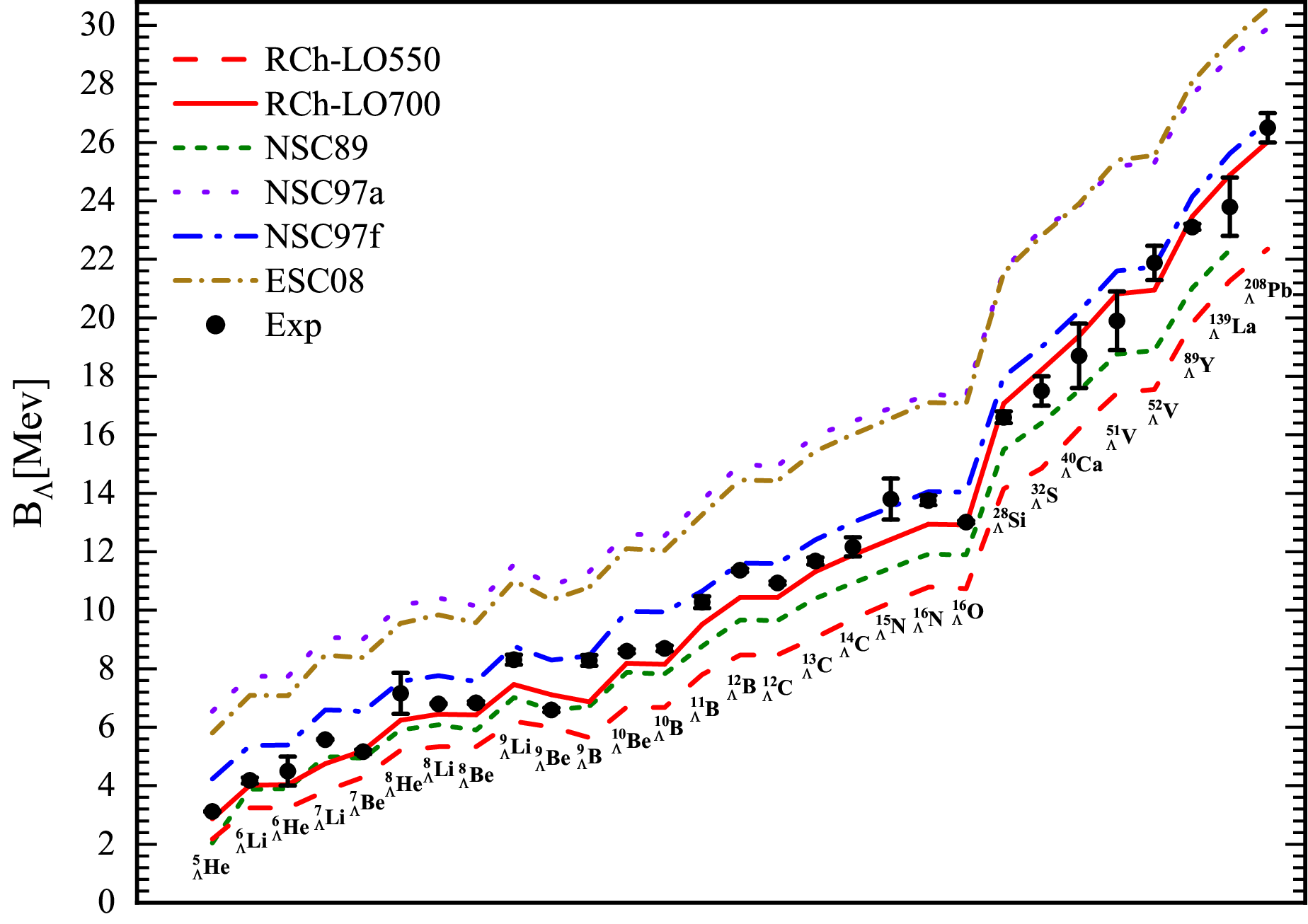}}
\vskip-3mm
\caption{
$\Lambda$ hyperon binding energies across light to heavy mass regions calculated using the interaction RCh-LO550, RCh-LO700, NSC89, NSC97a, NSC97f, and ESC08, compared with experimental data \cite{expbuilding1,expb2,expb3,expb4,expb5,expb6,expb7,expb9,expb10,expb11}. The red solid line represents the results obtained with our parameter set RCh-LO700, while the red dashed line corresponds to RCh-LO550. The green short-dashed line denotes NSC89, the purple dotted line indicates NSC97a, the blue dash-dotted line shows NSC97f, and the yellow short dot-dashed line represents ESC08.
}
\label{fig2}
\end{figure*}


\begin{table*}[t]
\centering
\caption{The RMS deviation $\Delta$ and the RRS deviation $\delta$ obtained with interactions RCh-LO550, RCh-LO700, NSC89, NSC97a, NSC97f, and ESC08, compared to experimental data. The RMS deviation is calculated as $\Delta = \sqrt{\sum_{i=1}^{N} (B_{\Lambda,i}^{\mathrm{Exp}} - B_{\Lambda,i}^{\mathrm{SHF}})^2 / N}$, and the RRS deviation is calculated as $\delta = \sqrt{\sum_{i=1}^{N} [(B_{\Lambda,i}^{\mathrm{Exp}} - B_{\Lambda,i}^{\mathrm{SHF}})/B_{\Lambda,i}^{\mathrm{Exp}}]^2 / N}$.}
\label{tablerms}
\begin{tabular}{l *{12}{c}}
\toprule
 & RCh-LO550 &\qquad \qquad \qquad & RCh-LO700 &\qquad \qquad \qquad & NSC89 &\qquad \qquad \qquad & NSC97a &\qquad \qquad \qquad & NSC97f &\qquad \qquad \qquad & ESC08 &\qquad \qquad \qquad \\
\midrule
$\Delta$\qquad \qquad \qquad  & 2.44 & & 0.71 & & 1.44 & & 4.00 & & 1.02 & & 5.35 & \\
$\delta$\qquad \qquad \qquad & 0.22 & & 0.08 & & 0.12 & & 0.47 & & 0.13 & & 0.40 & \\
\bottomrule
\end{tabular}
\end{table*}

\begin{figure*}[t]
\vskip-2mm
\centerline{\includegraphics[scale=0.40]{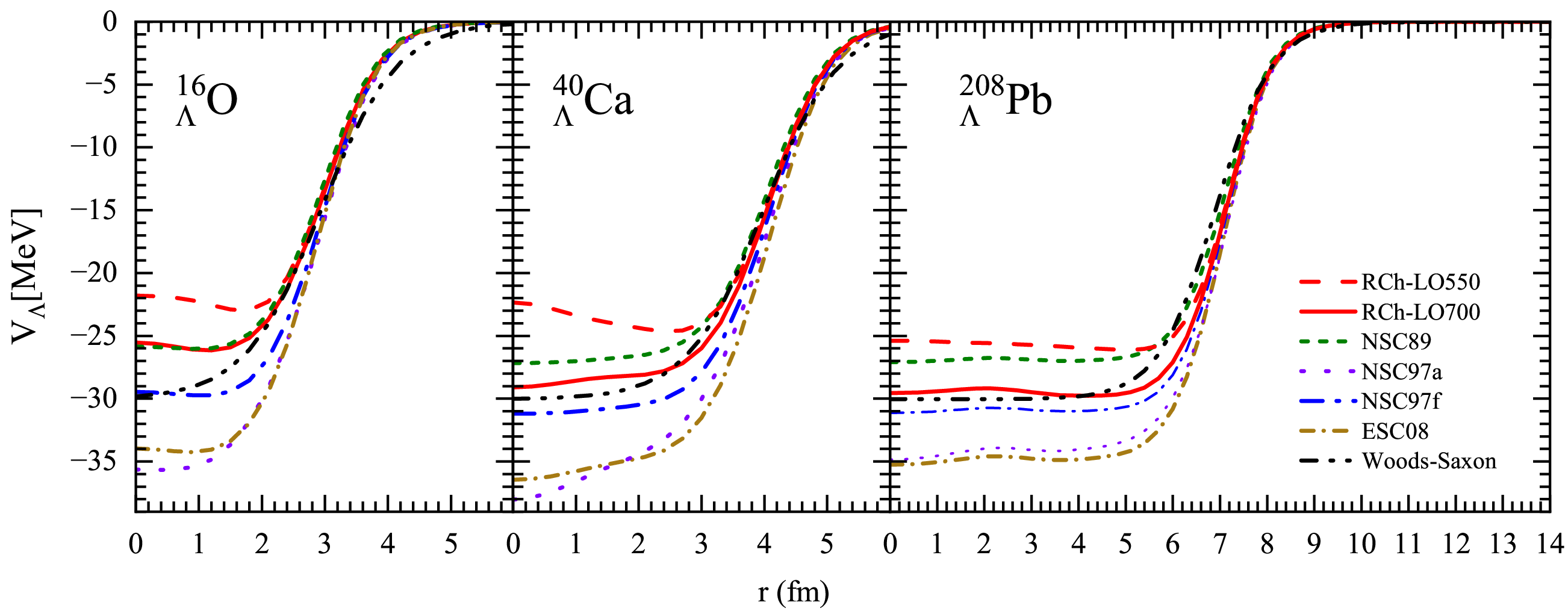}}
\vskip-3mm
\caption{
Local mean-field potentials for $^{16}_\Lambda\mathrm{O}$, $^{40}_\Lambda\mathrm{Ca}$, and $^{208}_\Lambda\mathrm{Pb}$ calculated with the microscopic interactions RCh-LO550, RCh-LO700, NSC89, NSC97a, NSC97f, and ESC08, along with the Woods-Saxon potential. The Woods-Saxon potential is taken from Ref.~\cite{gai2016} and is shown as a black dash-dotted line, while the line styles for the other interactions remain consistent with those in Fig.~\ref{fig2}.
}
\label{fig3}
\end{figure*}

\section{Results and discussion}
\label{sec3}

As explained in the preceding section, 
We have developed two parameter sets, RCh-LO550 and RCh-LO700, derived from the lower and upper cutoffs of the leading-order covariant chiral $YN$ potential, respectively. To initially evaluate their validity and characteristics, we begin by comparing these parameters with previously established microscopic interactions.

The effective mass $m_{\Lambda}^*/m_{\Lambda}$ and potential contributed by nucleons to hyperons $V_{\Lambda}$ calculated with different parameter sets are presented in Fig.~\ref{fig1}. Herein, we compare the RCh-LO550 and RCh-LO700 interactions 
with other microscopic interactions NSC89, NSC97a, NSC97f, and ESC08, all of which are derived from the Nijmegen soft-core potentials via BHF calculations \cite{hansnsc89,hansnsc97,hansesc08}.

It can be seen that the range of calculated effective mass for the parameters RCh-LO550 and RCh-LO700 we obtained is reasonable and shows no significant deviation when compared to other microscopic interactions.  
In RBHF, the effective mass is incorporated within the Dirac equation to account for the exchange of scalar mesons \cite{rbhf2}, whereas in BHF, it reflects the momentum dependence of particle kinetic energy as modified by the nuclear medium \cite{bhf2}.
Despite their different theoretical origins, the theoretical potentials represented by these interactions achieve a theoretically reliable level of accuracy in fitting scattering experimental data, and the nuclear matter calculation results from both RBHF and BHF are consistent with experimental observations to a certain extent \cite{nji3,nsc97,bare89,bare97,Rbare08,zheng2025}. Consequently, when these results of effective mass are incorporated into the SHF framework, the derived parameter sets, including our RCh-LO550 and RCh-LO700, naturally exhibit reasonable consistency with each other.

As the nucleon density increases, the difference between RCh-LO700 and RCh-LO550 in the effective mass becomes increasingly significant. Compared with RCh-LO700, the effective mass calculated with RCh-LO550 exhibits a steeper decline and soon falls below the value obtained with RCh-LO700. 
At the nucleon saturation density \( \rho_N = \rho_0 = 0.17\ \text{fm}^{-3} \), we observe that the \( m^*_\Lambda/m_\Lambda \) values of RCh-LO550 and RCh-LO700 are 0.750 and 0.642, respectively.  Among them, the \( m^*_\Lambda/m_\Lambda \)  obtained with RCh-LO700 is close to the empirical value of 0.780 reported in Ref.~\cite{gai2016}.

The $V_{\Lambda}$ can be obtained by taking the derivative of the $YN$ interaction density with respect to the hyperon density. For RCh-LO550 and RCh-LO700, similarly, as the nucleon density increases, the difference between the two parameter sets becomes more pronounced, and they even exhibit distinctly different behaviors at relatively high nucleon densities.

At saturation density, the potential depth of RCh-LO700 is also in close proximity to the empirical potential of \(V_{\Lambda} = -30\ \text{MeV}\) \cite{gai2016,gai2023}.
Meanwhile, we observe that at $\rho_{\Lambda}=0$, the potential well parameter behavior of RCh-LO700 closely resembles that of NSC97f. In the covariant chiral effective field theory based RBHF approach, the total single-particle potential obtained with cutoff = 700 MeV is close to that of NSC97f in the BHF framework. However, the repulsive contribution from the $^3P_1$ wave with cutoff = 700 MeV is slightly stronger than that of NSC97f \cite{zheng2025,nsc97}. Such results are favorable for the existence of high-mass neutron stars, while also leading to stronger suppression of $\Lambda$ hyperon clustering toward the core center in hypernuclei.

When the nucleon density exceeds \(0.16\ \text{fm}^{-3}\), the potential well depth of RCh-LO550 grows increasingly shallow with rising nucleon density, accompanied by a gradual weakening of its attractive force. This trend aligns with findings reported in Ref.~\cite{zheng2025}, which indicate that the attractive force intensifies as the cutoff increases. This characteristic exhibits a distinct trend compared to other microscopic interactions.
Such a characteristic is further corroborated by the $\Lambda$ hyperon binding energy results we present subsequently, RCh-LO550 predicts excessively weak hypernuclear binding in these calculations.

Single-particle $\Lambda$ hyperon binding energy in hypernuclei is the minimum energy required to remove a single hyperon from a hypernucleus
\begin{equation}
    B_{\Lambda}=E(^A_{\Lambda}Z)-E(^{A-1}_{\Lambda}Z).
\end{equation}
The experimentally measured binding energy results serve as an excellent constraint on $\Lambda N$ interaction models. More experimental constraints will help us rule out the inconsistent parts of the interaction theory.  
Since mean-field approximations are known to be less accurate for light nuclei \cite{shf1972,shf73,hansnsc89}, we performed calculations of $\Lambda$ binding energies for 29 hypernuclei, from light $^{5}_{\Lambda}\mathrm{He}$ to heavy $^{208}_{\Lambda}\mathrm{Pb}$, using the RCh-LO550 and RCh-LO700 interactions respectively. 
The $\Lambda$ hyperon has a weak perturbing effect on the overall deformation of the nuclear core \cite{zhou07,beChen2022,ChenCF22,guo22}. While the influence of deformation is relatively small for heavy hypernuclei as they remain nearly spherical, the correction due to deformation is non-negligible for the binding energy of light hypernuclei \cite{zhou07,ChenCF22}. Therefore, deformation has been  incorporated in our binding energy calculations for hypernuclei.
A comparison with other microscopic interactions NSC89, NSC97a, NSC97f, and ESC08 as well as with available experimental data is presented in Fig.~\ref{fig2}, and we employed the root-mean-square (RMS) deviation and the
relative square (RRS) deviation
to quantify the agreement between the calculated results and the experimental data. The results are presented in Table~\ref{tablerms}.

Results obtained with RCh-LO700, NSC89, and NSC97f show better agreement with the trend of experimental observations. 
Although the RCh-LO700 interaction exhibits some deviations in light hypernuclei, a behavior also shared by NSC89 and NSC97f, it overall achieves excellent agreement with experimental data, particularly in the heavy mass region. 
In the light mass region, RCh-LO700 provides a better description of $^8_\Lambda\text{Li}$ and $^8_\Lambda\text{Be}$ compared to NSC97f, whereas NSC97f more accurately reproduces the properties of $^9_\Lambda\text{Li}$ and $^9_\Lambda\text{B}$, a task in which RCh-LO700 performs poorly. Meanwhile, NSC89 offers the most accurate description for $^9_\Lambda\text{Be}$ among these interactions. Furthermore, none of these interactions can satisfactorily reproduce the experimental results for $^7_\Lambda\text{Li}$. These discrepancies may originate from limitations of the mean-field approximation, and physical effects, such as isospin effect that are not yet fully accounted for in these interactions~\cite{zhou07,CSB1,CSB2}. The light hypernuclear region thus remains an area rich in distinctive features worthy of further investigation.

As can see from Table~\ref{tablerms}, the RCh-LO700 interaction yields the smallest RMS value and RRS value among the compared microscopic interactions, including RCh-LO550 and those derived from the Nijmegen soft-core potentials. Given the relative scarcity of YN scattering data and the high accuracy of the bare potentials of these interactions in fitting the available YN scattering data, the LECs have a well-defined meaning in quantum field theory compared to the Nijmegen phenomenological parameters and exhibit lower model dependence than phenomenological models \cite{yn1,yn2,yn3,yn4,yn5,bare89,nsc97}. In nuclear matter, the leading-order covariant chiral effective potential with a cutoff of 700 MeV accurately describes the empirical single-particle potential. When applied to finite hypernuclei, it also successfully reproduces the binding energies of hypernuclei.The SHF microscopic interaction parameter set RCh-LO700, 
demonstrates overall excellent performance in predicting hyperon binding energies.

The binding energies obtained with the RCh-LO550 interaction are systematically smaller than the experimental values, indicating that the hypernuclear systems are weakly bound compared to experiment. This underbinding is linked to the weaker attraction provided by the in-medium single-particle potential of the 550 MeV cutoff. Although the stronger repulsion in the single-particle potential compared to non-relativistic LO and meson-exchange NSC97f interactions is promising for resolving the hyperon puzzle in neutron stars \cite{zheng2025}, the RCh-LO550 interaction may exhibit insufficient attraction for describing binding energies in finite hypernuclei. This result represents an opposite extreme when compared to the overbinding phenomena predicted by the NSC97a and ESC08 interactions. In BHF calculations, both NSC97a and ESC08 produce single-particle potentials that are overbound compared to empirical values. These interactions exhibit strong attraction in the $^3S_1$ wave and weak repulsion in the $^3P_1$ wave, a trend that is consistently reflected in hypernuclear binding energy calculations \cite{Rbare08,nsc97,bare97,nji3}.

\begin{figure}[t]
\centerline{\includegraphics[scale=0.24]{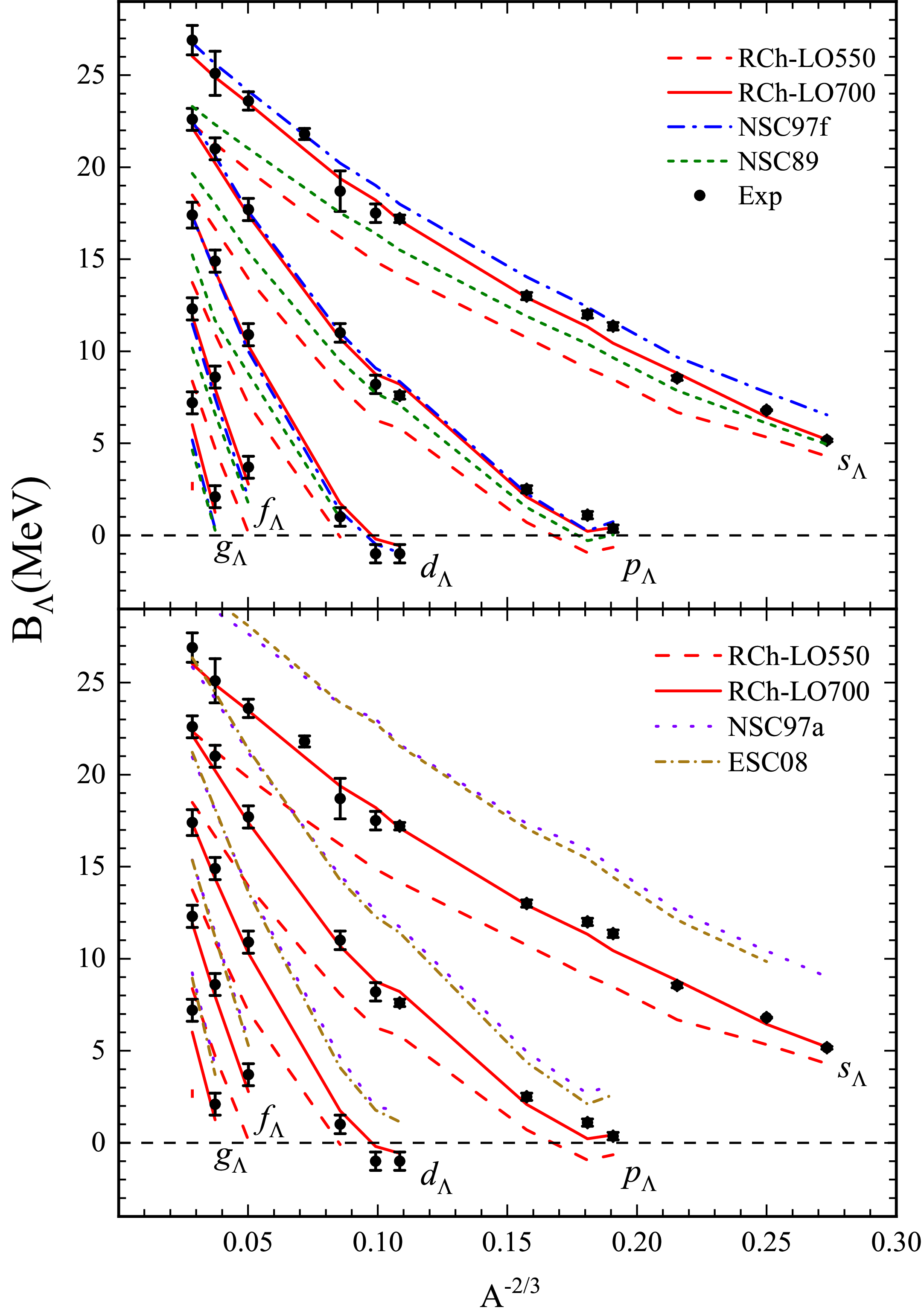}}
\vskip-2mm
\caption{
Energy levels of the $\Lambda$ single-particle major shells in $^{A}_{\Lambda}Z$ hypernuclei as a function of ${A}^{-2/3}$, calculated using the interactions RCh-LO550, RCh-LO700, NSC89, NSC97a, NSC97f, and ESC08, with experimental results included for comparison \cite{gai2016}. The line styles remain consistent with those in Fig.~\ref{fig3}.}
\label{fig4}
\end{figure}

Fig.~\ref{fig3} shows the mean-field profiles for the hypernuclei $^{16}_{\Lambda}\mathrm{O}$, $^{40}_{\Lambda}\mathrm{Ca}$, and $^{208}_{\Lambda}\mathrm{Pb}$ as calculated with different interactions. These results provide valuable insight for interpreting the computed binding energies of hypernuclei. Different parameter sets yield distinct potential well depths and widths. We also include for comparison the results from the empirical Woods-Saxon potential \cite{gai2016}. Among the hypernuclei $^{16}_{\Lambda}\mathrm{O}$, $^{40}_{\Lambda}\mathrm{Ca}$, and $^{208}_{\Lambda}\mathrm{Pb}$, the depth of the mean field generated by RCh-LO700 in $^{16}_{\Lambda}\mathrm{O}$ is very close to that of NSC89, with values of 25.54 MeV and 25.86 MeV, respectively.
 However, in heavy hypernuclei such as $^{208}_{\Lambda}\mathrm{Pb}$, the depth of the RCh-LO700 mean field decreases significantly, approaching the result of the Woods-Saxon potential, while the width of the potential well is slightly broader. The central potential value of $-29.55 \text{MeV}$ in $^{208}_{\Lambda}\mathrm{Pb}$ is consistent with the typical phenomenological potentials of Ref.~\cite{gai2023}. The notably larger variation in the depth of the RCh-LO700 central potential well from $^{16}_\Lambda\mathrm{O}$ to $^{208}_\Lambda\mathrm{Pb}$ compared to other interactions may contribute to its overall excellent performance in describing hypernuclear binding energies across the nuclear chart. Therefore, an appropriate hyperon mean-field potential is crucial for obtaining binding energies that are in close agreement with experimental observations.

The RCh-LO550 interaction produces the shallowest mean-field potential. Notably, at the center of $^{16}_{\Lambda}\mathrm{O}$, the depth of the mean-field generated by RCh-LO550 is only $-21.81 \text { MeV}$, which is significantly shallower than that of the empirical Woods-Saxon potential. This shallow potential provides a direct explanation for the underbinding of hypernuclear systems calculated with RCh-LO550. The NSC97a and ESC08 interactions produce the deepest potential wells in $^{16}_\Lambda\mathrm{O}$, $^{40}_\Lambda\mathrm{Ca}$, and $^{208}_\Lambda\mathrm{Pb}$; in particular, the depths of their central potential wells in $^{16}\mathrm{O}$ already reach 35.65 MeV and 33.98 MeV, respectively, and these wells provide stronger binding for the hypernuclear systems.

It is noteworthy that the mean-field potentials generated by these microscopic interactions almost coincide at large radii. This behavior can be attributed to the neglect of finite-range effects or gradient corrections in the SHF model \cite{beChen2022}.

Fig.~\ref{fig4} presents the binding energies for hyperons placed in different orbitals, computed using the RCh-LO550 and RCh-LO700 interactions, along with comparisons to other microscopic interactions and experimental data.

It should be noted that in deformed hypernuclei, energy levels of hyperon orbitals, with the exception of the s-orbital, are not degenerate. Therefore, in our calculations, the hyperon is placed in the lowest-energy state within a given excited orbital. Taking the 1p orbital as an example, the hyperon is assigned to the lowest single-particle energy level within the 1p orbital manifold.

Both RCh-LO700 and NSC97f yield results that agree well with experimental data. Around $\mathrm{A}^{-2/3} = 0.1$, the unbound experimental results for hyperons in the $d$ orbital are accurately reproduced by both the RCh-LO700 and NSC97f interactions. It is noteworthy that in lower orbitals, such as the $s$ and $p$ orbits, NSC97f produces stronger binding than RCh-LO700. Near $A^{-2/3} = 0.1$ in the $p$ orbit, the stronger binding exhibited by RCh-LO700 leads to better agreement with experimental values compared to NSC97f. In higher orbitals, such as the $d$, $f$, and $g$ orbits, RCh-LO700 conversely shows stronger binding than NSC97f, resulting in superior consistency with the two experimental data points in the $g$ orbit. 
It can be observed that the microscopic interaction RCh-LO700 provides an accurate description of experimental data even when the hyperon occupies different orbitals.

While the result obtained by RCh-LO550 continues to exhibit underbinding, a bound state can no longer be produced by this interaction when the hyperon is placed in the $g$ orbital for a mass number value of $\mathrm{A}^{-2/3} = 0.04$. In contrast, NSC97a and ESC08 show overbinding, and neither interaction can reproduce the unbound states of hyperons in the $d$ orbital near $\mathrm{A}^{-2/3} = 0.1$ that have been observed experimentally. The NSC89 interaction yields slightly underbound results,and this underbinding leads to poor descriptions of nuclei with larger mass numbers, regardless of which orbital the hyperon occupies.

\section{Summary}
\label{sec4}

We have constructed two sets of microscopic interactions RCh-LO550 and RCh-LO700 without any adjustable parameters, which are applicable to the SHF framework. These interactions are derived from RBHF calculations of nuclear matter using the leading-order covariant chiral $YN$ interaction, corresponding to the two potential sets with cutoffs of 550 MeV and 700 MeV, respectively. In finite hypernuclear systems, the effectiveness of the leading-order covariant chiral $YN$ interaction has been further validated. We calculated the binding energies of 29 hypernuclei and compared them with experimental values. The results show that the parameter set RCh-LO700 not only provides an excellent description of hypernuclear properties across the nuclear chart but also exhibits overall better performance than previous microscopic interactions derived from non-relativistic BHF calculations based on Nijmegen potentials. In contrast, the binding energies of finite hypernuclei calculated with the parameter set RCh-LO550 exhibit an underbinding phenomenon compared to experimental values.

The leading-order covariant chiral $YN$ force exhibits rigorously implemented Lorentz covariance and chiral symmetry, requires minimal higher-order corrections, and offers controllable regularization. This interaction not only enables the self-consistent description of experimentally consistent scattering cross-sections and empirical single-particle potentials but also has been validated in the present work to provide a robust description of experimental results for finite hypernuclei. Meanwhile, RCh-LO700, the newly constructed set of microscopic interactions, yields promising outcomes in hypernuclear calculations within the SHF framework.

\section*{Acknowledgments}

This work was supported by the National Natural Science
Foundation of China under Grants No. 12175071, No.12435007, No.12405133, No.12347180, and by the Fundamental Research Funds for the Central
Universities No.12575124.


\newcommand{\epja}{EPJA}
\newcommand{\npa}{Nucl. Phys. A}
\newcommand{\nphysa}{Nucl. Phys. A}
\newcommand{\ppnp}{Prog. Part. Nucl. Phys.}
\newcommand{\ptp}{Prog. Theor. Phys.}
\newcommand{\ptep}{Prog. Theor. Exp. Phys.}

\bibliography{rchlo}
\end{document}